\documentclass[sigconf]{acmart}
\AtBeginDocument{%
  }

\copyrightyear{2027}
\acmYear{2027}
\setcopyright{cc}
\setcctype{by}
\acmConference[KDD '27]{Proceedings of the 33rd ACM SIGKDD Conference on Knowledge Discovery and Data Mining V.1}{August 01--05, 2027}{San Jose, CA, USA}
\acmBooktitle{Proceedings of the 33rd ACM SIGKDD Conference on Knowledge Discovery and Data Mining V.1 (KDD '27), August 01--05, 2027, San Jose, CA, USA}
\acmDOI{10.1145/3862868.3871302}
\acmISBN{979-8-4007-3134-1/2027/08}

\usepackage{subcaption}
\usepackage{multirow}
\usepackage{enumitem}
\usepackage{amssymb}
\usepackage{pifont}

\usepackage{algorithm}
\usepackage{amsfonts}

\usepackage{listings}
\usepackage{url}

\usepackage{colortbl}  

\usepackage{xspace}

\newcommand{\best}[1]{\textbf{#1}}
\newcommand{\second}[1]{\underline{#1}}

\lstdefinestyle{custom_style}{
    basicstyle=\ttfamily\small,     
    breaklines=true,                 
    frame=single,                    
    xleftmargin=1pt,               
    xrightmargin=1pt,               
    keywordstyle=\color{blue},       
    commentstyle=\color{gray},      
    stringstyle=\color{red},         
}
\usepackage[capitalize]{cleveref}

\begin{document}


\title{Escaping the Euclidean Void: Manifold-Informed Flow Matching for Sequential Recommendation}

\author{Dengzhao Fang}
\affiliation{%
  \institution{Jilin University}
    \city{Changchun}
  \country{China}
}
\email{fangdz25@mails.jlu.edu.cn}

\author{Jingtong Gao}
\affiliation{%
  \institution{City University of Hong Kong}
    \city{Hong Kong}
  \country{China}
}
\email{jt.g@my.cityu.edu.hk}

\author{Yu Li}
\authornote{Corresponding author.}
\affiliation{%
  \institution{Jilin University}
    \city{Changchun}
  \country{China}
}
\email{liyu90@jlu.edu.cn}

\author{Xiangyu Zhao}
\affiliation{%
  \institution{City University of Hong Kong}
    \city{Hong Kong}
  \country{China}
}
\email{xianzhao@cityu.edu.hk}

\author{Yi Chang}
\affiliation{%
  \institution{Jilin University}
    \city{Changchun}
  \country{China}
}
\email{yichang@jlu.edu.cn}

\renewcommand{\shortauthors}{Fang et al.}

\begin{abstract}

Conventional recommenders capture users' preferences by optimizing user-item relations, 
whereas continuous generative recommendation reframes this process as the dynamic trajectory of generating a target representation.
While diffusion models build this trajectory through slow iterative denoising, 
flow matching accelerates generation by transporting source noise along straight probability paths in a continuous embedding space.
However, since item catalogs are inherently discrete and sparsely distributed, 
even these straight Euclidean paths often traverse unsupported regions devoid of valid item semantics.
Formalizing this failure as the \emph{Euclidean void}, we propose \textbf{MIRAGE}, 
a \textbf{M}anifold-\textbf{I}nformed
\textbf{R}ectification framework for \textbf{A}ccelerated \textbf{G}eneration of
\textbf{E}mbeddings in sequential recommendation, 
which rectifies the learned embedding geometry around an unchanged straight probability path.
By leveraging an item co-occurrence graph as a proxy for the underlying semantic manifold, MIRAGE aligns interpolated path states with local anchors, reorganizing the embedding space to ground the trajectory in valid item support.
MIRAGE retains the original probability path and uses the graph only during training, thereby enabling accurate and efficient one-step inference. 
Extensive experiments on four real-world datasets reveal that MIRAGE consistently outperforms state-of-the-art baselines, effectively boosting performance on sparsely observed targets while achieving robust overall accuracy.
Our code is available at \url{https://github.com/Applied-Machine-Learning-Lab/KDD2027_MIRAGE}. 
\end{abstract}



\maketitle

\section{Introduction}
\label{sec:introduction}

Sequential recommendation (SR) predicts a user's next interaction from an ordered behavior history, whose encoding has been progressively strengthened by recurrent, convolutional, Transformer, and LLM-based architectures~\cite{hidasi2016gru4rec,tang2018caser,kang2018sasrec,sun2019bert4rec,liu2025llmemb}. Most established objectives remain focused on target discrimination, as SASRec~\cite{kang2018sasrec} scores candidate items against a sequence representation, LightGCN~\cite{he2020lightgcn} scores items after propagating relational signals, and ranking or contrastive objectives make the observed target more compatible with the user than competing items. This design is effective because recommendation depends on the relative scores of these isolated discrete learned points and therefore does not require the interpolated continuous region between a user and target to carry meaningful item semantics.

Continuous generative recommendation changes this requirement because the model must learn the dynamic evolution of user intent rather than merely a target score, 
with diffusion-based methods perturbing target embeddings and recovering them through iterative denoising~\cite{ho2020ddpm,song2021score,yang2023dreamrec,li2024diffurec,li2025dimerec,chen2025unlocking}. 
Flow matching instead learns a time-dependent vector field and integrates an ordinary differential equation (ODE)~\cite{lipman2022flow}. 
Since this field is trained at intermediate states and queried along the generated path, its success heavily depends on the geometry encountered between the source noise and the target item. While target discrimination remains necessary for ranking, it is no longer sufficient for learning reliable continuous transport toward the precise target semantics.

Rectified flow improves transport efficiency by learning nearly straight ODE trajectories, which incur less discretization error and can be integrated with fewer solver steps~\cite{liu2022flow}. Following this straight-trajectory principle, FMRec~\cite{liu2025fmrec} adapts target-recovery flow matching and ODE sampling to SR, while FAVE~\cite{shi2026fave} further enables one-step transport through a history-informed prior and average-velocity modeling. However, while these methods optimize transport efficiency, they do not guarantee that the resulting straight path traverses meaningful regions of the embedding space. Consequently, a trajectory may be numerically simple yet semantically unsafe, causing the generated representations to drift into empty regions that yield inaccurate or irrelevant recommendations.

The source of this risk lies in the fundamental mismatch between a continuous Euclidean space and a discrete item catalog, where learned embeddings populate the space sparsely and unevenly. This aligns with the broader observation that high-dimensional data typically concentrates near lower-dimensional manifolds~\cite{bengio2013representation, fefferman2016manifold}. However, a straight interpolation path between a source and a target is not guaranteed to remain on this manifold~\cite{arvanitidis2018latent,chen2024rfm}. Consequently, an intermediate state can drift far from any valid item semantics, falling into what we term the \emph{Euclidean void}.

\begin{figure}[!t]
    \centering
    \includegraphics[width=\columnwidth]{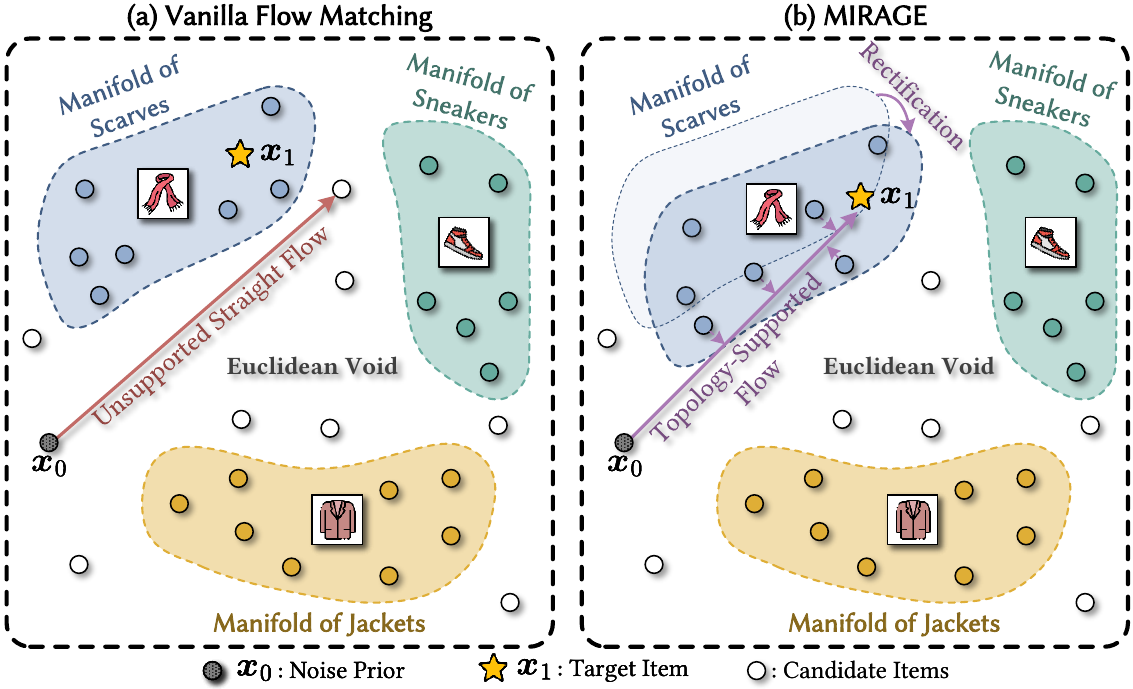}
    \caption{Euclidean void and MIRAGE rectification (illustration). (a) Vanilla straight paths traverse unsupported regions. (b) MIRAGE reorganizes item embeddings around the path via topology regularization to promote semantic support.}
    \Description{Two item-embedding spaces contain sparse candidate items and category-shaped regions for scarves, sneakers, and jackets. In the left panel, a red straight path from Gaussian noise to a scarf target crosses an unsupported region. In the right panel, MIRAGE retains a straight path but expands target-local scarf support around its intermediate states through training-time topology regularization.}
    \label{fig:intro}
\end{figure}

Unlike ODE discretization errors, the Euclidean void degrades the learning signal. Flow matching requires the neural network to infer the target direction from sampled states along the path. When a state falls into the void, it lacks surrounding geometric evidence, forcing the predictor to rely almost entirely on the user history to recover precise target semantics. 
Consequently, increasing the number of solver steps cannot resolve this structural failure, as finer numerical integration cannot synthesize missing catalog support.
The central challenge is therefore not only to learn a short path, but also to make the space traversed by that path semantically informative. This challenge is particularly critical for long-tail items,  whose representations receive limited direct supervision from user histories~\cite{kim2023melt,mao2025diqdiff,wei2023meta,wang2026sage}, relying on the local support of semantic neighbors to guide the trajectory away from the void.

Figure~\ref{fig:intro} makes this distinction concrete. A vanilla flow model projects a straight path toward the target, but crossing the Euclidean void causes the generated state to deviate from the true target. In contrast, our proposed framework retains the straight probability path while reorganizing valid item embeddings around its interior, ensuring the generated trajectory is semantically grounded and safely reaches the target. This contrast captures a shift from target-focused optimization to path-aware representation learning.

To realize this shift, we introduce \textbf{MIRAGE}, a \textbf{M}anifold-\textbf{I}nformed \textbf{R}ectification framework for \textbf{A}ccelerated \textbf{G}eneration of \textbf{E}mbeddings in SR. 
Since strict continuous manifolds are absent in discrete item catalogs, addressing the Euclidean void requires establishing semantic support along the straight probability trajectory. 
We treat an item co-occurrence graph as a discrete proxy for the underlying data manifold, exploiting the strong transition regularities inherent in user histories~\cite{barkan2016item2vec, wu2019srgnn}. 
Using this proxy, MIRAGE applies a topology regularizer that aligns the interpolated path states with active graph anchors drawn from the target's neighborhood.

Applying this regularization uniformly, however, risks corrupting the source noise and target representations at the endpoints. We resolve this boundary-preservation challenge through a dynamic time-modulation schedule that suppresses regularization at both extremes, focusing spatial restructuring on the unsupported path interior. This selectively repositions locally related embeddings to form a cohesive semantic manifold enveloping the generative trajectories. Furthermore, to avoid the high latency of inference-time graph lookups, MIRAGE confines topology supervision to the training phase. The co-occurrence graph anchors intermediate states to rectify the shared embedding space only during optimization. Since this geometric reorganization is absorbed directly into the item representations, the space itself provides a well-conditioned transport route. Consequently, inference operates without auxiliary graph traversals, keeping generated trajectories semantically grounded while preserving the computational efficiency of graph-free, one-step deterministic transport.

The main contributions of this work are summarized as follows:
\begin{itemize}[leftmargin=*,nosep]
    \item We introduce a new perspective on continuous generative recommendation by identifying locally supported intermediate states as a critical prerequisite for reliable transport and formalizing the absence of this semantic grounding as the \emph{Euclidean void}.
    \item We propose MIRAGE, a novel manifold-informed rectification framework that treats discrete co-occurrence graphs as proxy manifolds to dynamically reorganize the underlying embedding geometry around straight transport paths.
    \item We develop a trajectory-aware geometric rectification mechanism with boundary-preserving temporal modulation, which aligns intermediate states with local semantic support while maintaining the original flow formulation.
    \item Comprehensive evaluations across four datasets validate MIRAGE's superiority in state-of-the-art overall accuracy, robust long-tail gains, and one-step inference efficiency.
\end{itemize}

\section{Related Work}
\label{sec:related}

\subsection{Generative Sequential Recommendation}
Sequential recommendation (SR) models encode user behavior through deep sequence models (e.g., recurrent~\cite{hidasi2016gru4rec}, convolutional~\cite{tang2018caser}, and self-attentive networks~\cite{kang2018sasrec,sun2019bert4rec}) or contrastive~\cite{qiu2022duorec}, variational~\cite{sachdeva2019svae}, and translation-based~\cite{he2017transrec} objectives. Graph-based variants propagate collaborative signals over session~\cite{wu2019srgnn} or interaction graphs~\cite{he2020lightgcn,ye2023gmae}, while tail-enhancement frameworks transfer information to sparsely observed targets~\cite{kim2023melt}. 
More recently, LLM-driven and multi-agent paradigms have broadened semantic content
processing~\cite{liu2025llmemb,fang2025collaborative,fang2026llm4cgds} and interactive
agentic search~\cite{he2026localsearchbench,he2026vistahop}.
While these methods establish the predictive value of relational and semantic structure,
they primarily use it to encode static representations for direct target scoring.
This optimizes the latent space for endpoint similarity without requiring the continuous regions between users and items to carry meaningful semantics.

Moving beyond this static point estimation, generative SR spans discrete semantic-ID paradigms~\cite{fang2025hid} as well as diffusion-based methods that perturb target representations and recover them through iterative denoising~\cite{ho2020ddpm,song2021score}. Recent advancements have improved this diffusion process by generating ideal target representations~\cite{yang2023dreamrec}, modeling multiple interests~\cite{li2025dimerec}, mitigating data sparsity via diffusion-based augmentation~\cite{liu2023diffusion,ma2024plug}, addressing popularity bias and corruption stability~\cite{li2024diffurec,chen2025unlocking,mao2025diqdiff,bai2025bbdrec}, or incorporating item co-occurrence graphs to strengthen diffusion guidance~\cite{xie2026icusrec}. 
Although these diffusion-based methods successfully introduce generative conditioning, they collectively inherit discrete Markovian dynamics that require costly multi-step sampling and suffer from compounding discretization errors during inference; a stepwise drift is also observed across long-horizon agentic trajectories~\cite{he2026workrover,he2026checkerbench}.

\subsection{Flow Matching and Geometric Support}
To overcome the inefficiencies of discrete diffusion, flow matching (FM) regresses continuous vector fields along prescribed straight probability paths~\cite{lipman2022flow,liu2022flow,pooladian2023multisample,kornilov2024optimalflow}. In SR, FMRec pioneered this transition by adapting target-recovery FM with deterministic Euler sampling~\cite{liu2025fmrec}. 
Subsequent methods like FlowRec~\cite{li2025flowrec}, FAVE~\cite{shi2026fave}, FlowCF~\cite{liu2025flowcf}, and GMFlowRec~\cite{ye2026gaussian} accelerate generation by optimizing priors, velocities, or mixture distributions.
Although recent attempts like DAE4Rec~\cite{song2025enhancing} incorporate graph guidance for condition enhancement, these methods uniformly assume their continuous paths traverse semantically safe regions, largely overlooking the underlying geometric validity.

The vulnerability of these unsupported continuous paths is rooted in representation geometry. The manifold hypothesis posits that high-dimensional data concentrates near structured, lower-dimensional subsets~\cite{bengio2013representation,fefferman2016manifold}. Standard Euclidean interpolants often deviate from these curved manifolds, leading to regions devoid of meaningful semantics~\cite{arvanitidis2018latent,chen2024rfm}. While classical manifold regularization promotes smoothness among observed samples via graph structures~\cite{belkin2006manifold}, and Riemannian FM constructs explicit flows on known continuous manifolds~\cite{chen2024rfm}, generative recommendation lacks an explicit continuous manifold to guide the transport. Building upon this geometric perspective, we propose \textbf{MIRAGE} to resolve the Euclidean void. 
Rather than redesigning the transport velocity or requiring a known continuous manifold, MIRAGE leverages training-only discrete topology to rectify the shared embedding geometry around straight probability paths. This decouples spatial semantic safety from trajectory acceleration, ensuring that generative paths remain reliably grounded in valid representation geometry.

\section{Preliminaries}
\label{sec:preliminaries}

\subsection{Sequential Recommendation}

Let $\mathcal{U}$ and $\mathcal{I}$ denote the user set and the item catalog, respectively. For a given user $u\in\mathcal{U}$, the ordered history is $S_u=(i_1,\ldots,i_m)$ with $i_\ell\in\mathcal{I}$. Sequential recommendation ranks candidate items as the next interaction $i_{m+1}$, denoted by $i^+$. Each item $i$ is associated with a trainable embedding $\boldsymbol{e}_i\in\mathbb{R}^d$.

Generative sequential recommendation models depart from direct discrimination over the discrete item catalog, modeling a transport process in a continuous latent space to generate the target representation. Given a tractable source distribution $p_0$ (e.g., standard Gaussian) and a target distribution $p_1$ over next-item embeddings, the model learns to transport a source sample $\boldsymbol{x}_0 \sim p_0$ toward a target sample $\boldsymbol{x}_1 = \boldsymbol{e}_{i^+} \sim p_1$. Diffusion-based recommenders instantiate this paradigm through iterative denoising~\cite{yang2023dreamrec,li2024diffurec}, whereas flow-based recommenders learn deterministic probability paths~\cite{liu2025fmrec}.

\subsection{Flow Matching}

Flow matching learns a time-dependent vector field that transports samples from $p_0$ to $p_1$ through an ordinary differential equation (ODE)~\cite{lipman2022flow,liu2022flow}. Under the common linear probability path, a training pair $(\boldsymbol{x}_0, \boldsymbol{x}_1)$ defines an interpolated state
\begin{equation}
\label{eq:prelim_interp}
    \boldsymbol{x}_t = (1-t)\boldsymbol{x}_0 + t\boldsymbol{x}_1, \quad t \in [0,1].
\end{equation}
The corresponding conditional velocity is
\begin{equation}
\label{eq:prelim_velocity}
    \boldsymbol{u}_t\left(\boldsymbol{x}_t \;\middle|\; \boldsymbol{x}_0,\boldsymbol{x}_1\right)
    = \frac{d\boldsymbol{x}_t}{dt}
    = \boldsymbol{x}_1 - \boldsymbol{x}_0.
\end{equation}
In the sequential recommendation setting, conditional flow matching trains a neural vector field $\boldsymbol{v}_\theta$ conditioned on the user history $S_u$ to approximate this direction by minimizing the objective
\begin{equation}
\label{eq:prelim_cfm}
    \mathcal{L}_{\text{CFM}}
    =
    \mathbb{E}_{t,\boldsymbol{x}_0,\boldsymbol{x}_1, S_u}
    \left\|
    \boldsymbol{v}_\theta(\boldsymbol{x}_t,t,S_u)
    -
    \left(\boldsymbol{x}_1-\boldsymbol{x}_0\right)
    \right\|_2^2.
\end{equation}
Target-recovery flow matching instead directly generates the clean target embedding $\boldsymbol{x}_1$ and derives the velocity from that generation. MIRAGE adopts this parameterization in Section~\ref{sec:method}.

\subsection{Co-occurrence Neighborhoods}

Flow matching operates in a continuous latent space, but the item catalog is discrete. We construct an item graph $\mathcal{G}=(\mathcal{I},\mathcal{E})$ from training-set co-occurrences only. Edges count ordered item pairs co-occurring within a context window of size $c$, and $\mathcal{N}_K(i)$ contains the $K$ items with the largest nonzero counts for item $i$. The graph is precomputed and fixed during optimization.

The graph provides a criterion for measuring whether an interpolated state is close to the target-local co-occurrence neighborhood. For an interpolated state $\boldsymbol{x}_t$ associated with target item $i^+$, we define its graph-neighborhood distance as
\begin{equation}
\label{eq:prelim_knn_distance}
    d_{\mathcal{G}}(\boldsymbol{x}_t, i^+)
    =
    \min_{j \in \mathcal{N}_K(i^+)}
    \left\|\boldsymbol{x}_t - \boldsymbol{e}_j\right\|_2^2.
\end{equation}
A large value indicates $\boldsymbol{x}_t$ is far from the target's co-occurrence neighborhood despite being a valid Euclidean vector, thus quantifying the Euclidean void introduced in Section~\ref{sec:introduction}.

\section{Methodology}
\label{sec:method}

\subsection{Overview}

\begin{figure*}[!t]
    \centering
    \includegraphics[width=0.90\textwidth]{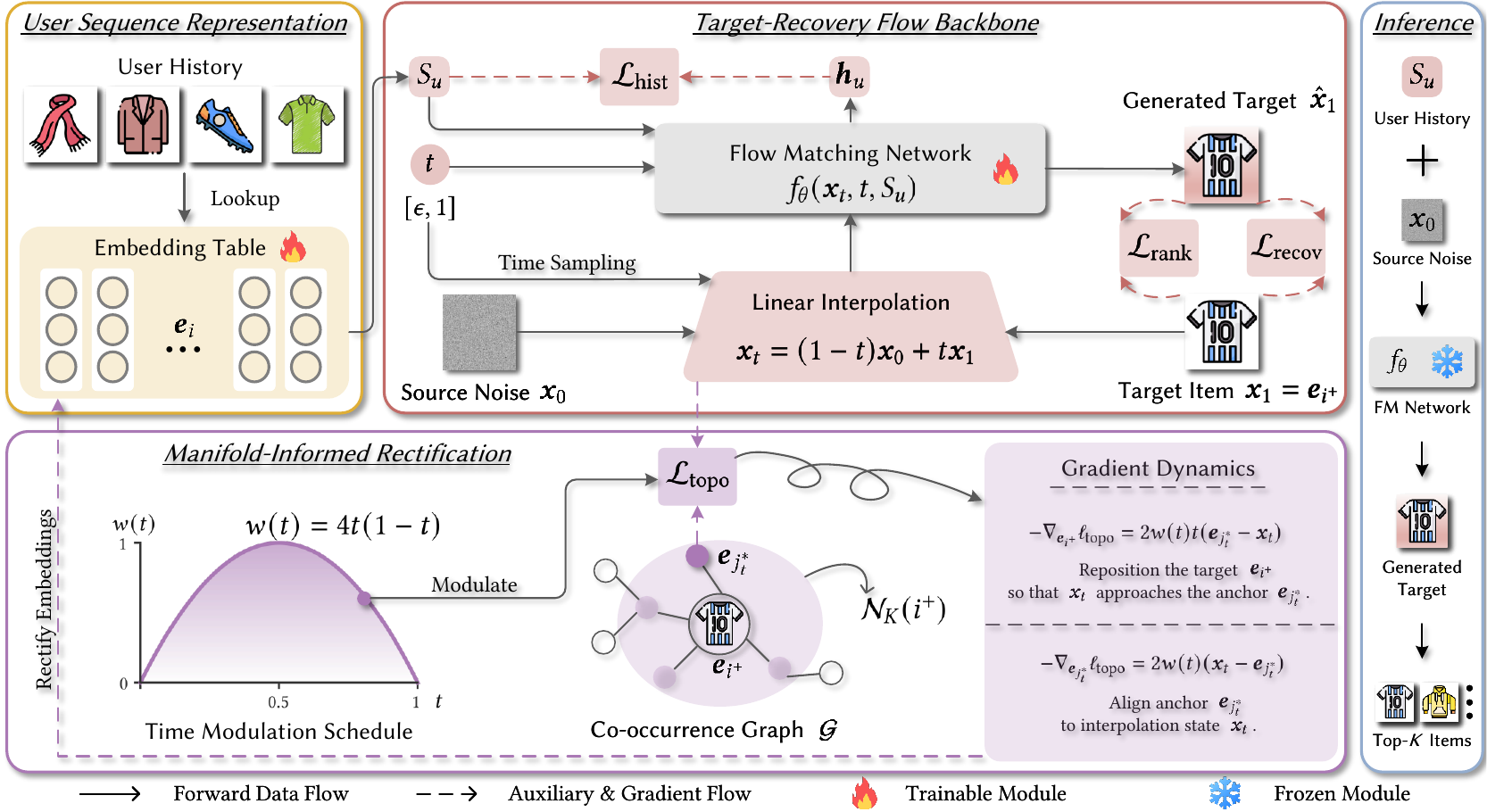}
    \caption{Overview of MIRAGE. The backbone $f_\theta$ predicts targets from interpolated user-conditioned states. A manifold-informed rectification mechanism dynamically reshapes the embedding space by aligning interpolated path states with local graph anchors, ensuring the generative path remains grounded in a valid semantic manifold.}
    \Description{Overview diagram with three connected modules. The upper-left module maps a user's item history through a trainable embedding table. The upper-right target-recovery backbone forms an interpolated state from source noise and the target embedding, then predicts the clean target conditioned on time and history while applying recovery, ranking, and history losses. The lower module represents the manifold-informed rectification mechanism, which combines a time modulation schedule, a target co-occurrence neighborhood, and a topology loss. Its joint gradient updates reposition the target and active anchor embeddings around the same linear interpolation.}
    \label{fig:method}
\end{figure*}

As illustrated in Figure~\ref{fig:method}, MIRAGE introduces a manifold-informed rectification mechanism over target-local co-occurrence neighborhoods. This dynamically reshapes the shared item geometry during training to encourage straight target-recovery trajectories to remain semantically supported while leaving inference graph-free. For a user history $S_u$, source noise $\boldsymbol{x}_0\sim\mathcal{N}(\boldsymbol{0},\boldsymbol{I})$, and target $\boldsymbol{x}_1=\boldsymbol{e}_{i^+}$, a flow matching network $f_\theta$ predicts the clean target $\hat{\boldsymbol{x}}_1$ directly from the interpolated state at time $t \in [\epsilon, 1]$.

\subsection{The Euclidean Void and Manifold Mismatch}
\label{subsec:euclidean_mismatch}

Building upon the conceptual intuition provided in Section~\ref{sec:introduction}, we now formalize the \emph{Euclidean void} as a measurable geometric mismatch inherent to continuous recommendation spaces. 
Standard flow matching optimizes the vector field to approximate the target direction by minimizing $\mathcal{L}_{\text{CFM}}$ in Eq.~\eqref{eq:prelim_cfm}. However, this objective strictly supervises the velocity field, 
leaving the spatial and semantic validity of the intermediate states $\boldsymbol{x}_t$ unconstrained.

Since the ideal linear interpolation path in Eq.~\eqref{eq:prelim_interp} is constructed in a global Euclidean space, it remains blind to the discrete underlying item structure. The exposure to the Euclidean void during training can be quantified by the expected geometric deviation along this prescribed probability path. Using the graph-neighborhood distance from Eq.~\eqref{eq:prelim_knn_distance}, the expected path deviation is defined as
\begin{equation}
\label{eq:path_deviation}
    \mathcal{D}_{\text{void}}(i^+) = \mathbb{E}_{t \sim U(0,1)} \left[ d_{\mathcal{G}}(\boldsymbol{x}_t, i^+) \right].
\end{equation}
For standard continuous recommenders, this expected deviation $\mathcal{D}_{\text{void}}(i^+)$ is usually large. 
The mismatch is most consequential in the path interior, where the model must infer target-directed transport without the semantic certainty available at the endpoints, easily drifting into unsupported continuous regions.

Eq.~\eqref{eq:path_deviation} translates the Euclidean void into a quantifiable training proxy. In standard models, the generator is forced to decode precise target semantics from intermediate vectors lacking local semantic support. To ensure that the actual generative trajectories remain reliable, the underlying space must actively suppress $\mathcal{D}_{\text{void}}(i^+)$. Rather than computing an explicit continuous manifold, MIRAGE treats this prescribed deviation as a direct topology regularizer, letting the geometric penalties reshape the trainable item geometry to safely envelop generative trajectories.

\subsection{Target-Recovery Flow Backbone}
\label{subsec:flow_backbone}
We adopt a target-recovery backbone as the generative foundation, optimized for target recovery, item discrimination, and historical reconstruction. We detail these base objectives before integrating our core manifold-informed rectification.

\subsubsection{Target Recovery}
To align with how sequential recommendation retrieves items via inner products, MIRAGE is designed to generate the target embedding directly. 
Consequently, we parameterize the conditioned vector field $\boldsymbol{v}_\theta$ to approximate the ground-truth conditional velocity from Eq.~\eqref{eq:prelim_velocity} by predicting the target state:
\begin{equation}
\label{eq:method_velocity_param}
    \boldsymbol{v}_\theta(\boldsymbol{x}_t,t,S_u)
    =
    f_\theta(\boldsymbol{x}_t,t,S_u)-\boldsymbol{x}_0,
\end{equation}
where $f_\theta$ denotes the flow matching network. Substituting this parameterization into Eq.~\eqref{eq:prelim_cfm} gives
\begin{equation}
\label{eq:method_recovery}
\begin{aligned}
    \mathcal{L}_{\text{recov}}
    &=
    \mathbb{E}_{t,\boldsymbol{x}_0,\boldsymbol{x}_1, S_u}
    \left\|
    (f_\theta(\boldsymbol{x}_t,t,S_u)-\boldsymbol{x}_0)
    -
    (\boldsymbol{x}_1-\boldsymbol{x}_0)
    \right\|_2^2 \\
    &=
    \mathbb{E}_{t,\boldsymbol{x}_0,\boldsymbol{x}_1, S_u}
    \left\|
    \hat{\boldsymbol{x}}_1-\boldsymbol{x}_1
    \right\|_2^2,
\end{aligned}
\end{equation}
where $\hat{\boldsymbol{x}}_1 = f_\theta(\boldsymbol{x}_t,t,S_u)$ is the generated target state. By eliminating the source noise $\boldsymbol{x}_0$, this formulation shows that matching the conditional velocity is equivalent to minimizing the Euclidean distance to the clean target embedding $\boldsymbol{x}_1$.

\subsubsection{Catalog Ranking}
Recovery alone pulls generations toward target coordinates and can collapse trainable item embeddings~\cite{yang2023dreamrec,chen2025unlocking}. To enforce robust contrastive separation, we augment training with a standard catalog-ranking objective over either the full item set or a sampled candidate set $\mathcal{C}_u$ that includes $i^+$. The objective is
\begin{equation}
\label{eq:method_rank_loss}
    \mathcal{L}_{\text{rank}}
    =-\log
    \frac{\exp(\hat{\boldsymbol{x}}_1^\top \boldsymbol{e}_{i^+})}
    {\sum_{a \in \mathcal{C}_u}
    \exp(\hat{\boldsymbol{x}}_1^\top \boldsymbol{e}_a)}.
\end{equation}
This auxiliary term preserves catalog-level discrimination while $\mathcal{L}_{\text{recov}}$ learns the continuous transport map.

\subsubsection{History Reconstruction}
To preserve historical preference semantics under continuous noise perturbations, we incorporate a history reconstruction objective~\cite{liu2025fmrec,shi2026fave}. 
Specifically, after the flow backbone encodes the user sequence into a final hidden state $\boldsymbol{h}_u$, we train a lightweight decoder $g_\omega$ to reconstruct the multi-hot user history vector 
$\boldsymbol{r}_u \in \{0,1\}^{|\mathcal{I}|+1}$
directly from this representation, yielding the predicted
interaction scores
$\hat{\boldsymbol{r}}_u = g_\omega(\boldsymbol{h}_u)$.

To maintain efficiency, the loss is evaluated over a targeted subset $\Omega_u$, which consists of all observed items and a sampled set of unobserved items. The reconstruction objective is a masked mean squared
error normalized by the full item vocabulary size:
\begin{equation}
\label{eq:method_hist_loss}
    \mathcal{L}_{\text{hist}}
    =
    \frac{1}{|\mathcal{B}|(|\mathcal{I}|+1)}
    \sum_{u\in\mathcal{B}}
    \sum_{a\in\Omega_u}
    \left(\hat{r}_{u,a}-r_{u,a}\right)^2,
\end{equation}
where $\mathcal{B}$ denotes a mini-batch of users, while $\hat{r}_{u,a}$ and $r_{u,a}$ denote the scalar values of the predicted vector $\hat{\boldsymbol{r}}_u$ and the ground-truth vector $\boldsymbol{r}_u$ at index $a$, respectively.

\subsection{Manifold-Informed Rectification}
\label{subsec:topology_regularization}

To implement the manifold-informed rectification, MIRAGE overcomes the Euclidean void by anchoring each interpolated state to its target-local neighborhood. The resulting gradients dynamically reshape the embedding space around the unchanged probability path. We formulate this through a time-modulated topology regularizer.

\subsubsection{Target-Local Alignment}
To explicitly enforce spatial validity, MIRAGE relies on \emph{hard alignment}, which anchors an interpolated state to a single valid item in the target-local neighborhood. Recalling Eq.~\eqref{eq:prelim_knn_distance} and assuming a unique nearest neighbor, we define the active anchor $j_t^*$ and its gradient:
\begin{equation}
\label{eq:method_anchor_gradient}
    j_t^* = \underset{j\in\mathcal{N}_K(i^+)}{\arg\min} \|\boldsymbol{x}_t-\boldsymbol{e}_j\|_2^2,
    \ \ \ \;
    \nabla_{\boldsymbol{x}_t} d_{\mathcal{G}}(\boldsymbol{x}_t,i^+) = 2(\boldsymbol{x}_t-\boldsymbol{e}_{j_t^*}).
\end{equation}
This unique minimum renders the distance function differentiable almost everywhere, yielding a gradient that pulls the interpolant toward the semantic anchor $\boldsymbol{e}_{j_t^*}$. 
Alternatively, we formulate a \emph{soft alignment} variant
with temperature-weighted local neighbors:
\begin{equation}
\begin{aligned}
    d_{\mathcal{G},\tau}^{\text{soft}}
    &=
    \sum_{j\in\mathcal{N}_K(i^+)}
    \alpha_{t,j}
    \|\boldsymbol{x}_t-\boldsymbol{e}_j\|_2^2,\\
    \alpha_{t,j}
    &=
    \frac{
    \exp(-\|\boldsymbol{x}_t-\boldsymbol{e}_j\|_2^2/\tau)
    }{
    \sum_{a\in\mathcal{N}_K(i^+)}
    \exp(-\|\boldsymbol{x}_t-\boldsymbol{e}_a\|_2^2/\tau)
    }.
\end{aligned}
\end{equation}
The resulting objective smoothly weights neighboring item distances according to their proximity to the interpolated state.
Unlike standard graph-based contrastive learning where continuous averages of discrete items may fall into unsupported regions, hard alignment offers a stricter geometric constraint on the discrete item manifold while soft alignment provides a smoother optimization landscape.
We evaluate both variants empirically, providing complete theoretical derivations in Appendix~\ref{app:alignment-derivation}.

\subsubsection{Time Modulation Schedule}
To enforce the aforementioned spatial alignment without violating the boundary conditions of the generative flow, MIRAGE applies this target-local constraint dynamically. The regularizer must strictly vanish at $t=0$ to preserve the unstructured source prior, and at $t=1$ to leave the clean target state unconstrained. Among symmetric quadratic schedules normalized to a peak of $w(1/2)=1$, these standard boundary conditions yield:
\begin{equation}
\label{eq:method_time_weight}
    w(t)=4t(1-t).
\end{equation}
This schedule concentrates semantic guidance in the path interior, where the interpolant mixes the source and target strongly. Its parabolic shape matches the uncertainty profile $t(1-t)$ of a unit Brownian bridge, serving as a geometric analogy~\cite{karatzas1991brownian}. By applying this weight to minimize the expected path deviation $\mathcal{D}_{\text{void}}(i^+)$ defined in Eq.~\eqref{eq:path_deviation}, we formulate the target-local regularizer:
\begin{equation}
\label{eq:method_topo_loss}
    \mathcal{L}_{\text{topo}}
    =
    \mathbb{E}_{t,\boldsymbol{x}_0,\boldsymbol{x}_1, S_u}
    \left
    [
    w(t) \cdot \|\boldsymbol{x}_t-\boldsymbol{e}_{j_t^*}\|_2^2
    \right
    ].
\end{equation}

\subsubsection{Gradient Dynamics and Geometric Rectification}
To analyze how this regularizer reshapes the space, consider its per-sample integrand $\ell_{\text{topo}}=w(t)\|\boldsymbol{x}_t-\boldsymbol{e}_{j_t^*}\|_2^2$. The active anchor is locally constant almost surely, and $\partial\boldsymbol{x}_t/\partial\boldsymbol{e}_{i^+}=t\boldsymbol{I}$. The chain rule yields:
\begin{equation}
\label{eq:method_embedding_gradients}
\begin{aligned}
    \nabla_{\boldsymbol{e}_{i^+}}\ell_{\text{topo}}
    &=2w(t)t\left(\boldsymbol{x}_t-\boldsymbol{e}_{j_t^*}\right),\\
    \nabla_{\boldsymbol{e}_{j_t^*}}\ell_{\text{topo}}
    &=-2w(t)\left(\boldsymbol{x}_t-\boldsymbol{e}_{j_t^*}\right).
\end{aligned}
\end{equation}
Through these updates, gradient descent repositions the target embedding $\boldsymbol{e}_{i^+}$ so that the interpolated state $\boldsymbol{x}_t$ approaches the anchor $\boldsymbol{e}_{j_t^*}$, while simultaneously aligning the anchor $\boldsymbol{e}_{j_t^*}$ toward the current interpolation state. The time-modulation factor $w(t)$ suppresses both updates at the source and target boundaries, concentrating the spatial reorganization exclusively in the path interior. Appendix~\ref{app:gradient-dynamics} demonstrates that these joint updates contract the residual distance under a small step size.

Notably, these pairwise forces update the shared embedding table without explicit graph message passing. Although the topology term does not directly update flow parameters, it reorganizes the Euclidean space to envelop sampled path interiors with valid item evidence. This reshapes the target trajectories into a smoother continuous vector field for the flow backbone to regress. 
Consequently, this path-interior topology supervision restructures the space into a well-behaved vector field, minimizing discretization errors and ensuring accurate one-step target recovery.

\subsection{Training Objective and Inference}
\label{subsec:objective_inference}

The full training objective integrates the flow backbone and the geometric regularization:
\begin{equation}
\label{eq:method_total_loss}
    \mathcal{L}_{\text{MIRAGE}}
    =
    \mathcal{L}_{\text{recov}}
    +
    \lambda_{\text{rank}}\mathcal{L}_{\text{rank}}
    +
    \lambda_{\text{hist}}\mathcal{L}_{\text{hist}}
    +
    \lambda_{\text{topo}}\mathcal{L}_{\text{topo}}.
\end{equation}

At inference time, MIRAGE operates without requiring any graph computations. Given a user history $S_u$, the model samples a source noise $\boldsymbol{x}_0 \sim \mathcal{N}(\boldsymbol{0}, \boldsymbol{I})$ and generates the target representation via $q$-step deterministic Euler sampling ($q \in \mathbb{N}^+$). 
Dividing the integration interval $[\epsilon, 1]$ into $q$ uniform steps, the network times are discretized as $t_n=\epsilon+(1-\epsilon)\frac{n}{q}$, for $n=0,\ldots,q-1$. Following the linear velocity formulation in Eq.~\eqref{eq:prelim_velocity}, our target-recovery network explicitly predicts the clean target $\boldsymbol{x}_1$ as $f_\theta(\tilde{\boldsymbol{x}}_n, t_n, S_u)$. Thus, the intermediate states are recursively updated via Euler integration using the exact step size $\Delta t = \frac{1-\epsilon}{q}$:
\begin{equation}
\label{eq:method_euler}
    \tilde{\boldsymbol{x}}_{n+1}
    =
    \tilde{\boldsymbol{x}}_n
    +
    \frac{1-\epsilon}{q}
    \left[
    f_\theta(\tilde{\boldsymbol{x}}_n,t_n,S_u)-\boldsymbol{x}_0
    \right],
\end{equation}
with $\tilde{\boldsymbol{x}}_0=\boldsymbol{x}_0$. 
For efficient one-step generation ($q=1$), this recursion evaluates to $\tilde{\boldsymbol{x}}_1 = (1-\epsilon)f_\theta(\boldsymbol{x}_0,\epsilon,S_u) + \epsilon \boldsymbol{x}_0$. Since the minimum network time $\epsilon$ is negligible, this expression reduces to
\begin{equation}
\label{eq:method_onestep}
    \hat{\boldsymbol{x}}_1=f_\theta(\boldsymbol{x}_0,\epsilon,S_u).
\end{equation}
Setting the final target representation as $\hat{\boldsymbol{x}}_1 \equiv \tilde{\boldsymbol{x}}_q$, we compute the candidate scores for ranking via inner products:
\begin{equation}
\label{eq:method_scores}
    s_a=\hat{\boldsymbol{x}}_1^\top \boldsymbol{e}_a,\quad \forall a \in \mathcal{I}.
\end{equation}
The integration step $q$ balances inference cost and refinement. Our training-time topological guidance enables zero-graph overhead during generation while maintaining robust recommendation quality even at $q=1$ as demonstrated in Section~\ref{subsec:efficiency}.

\section{Experiments}
\label{sec:exp}

We evaluate MIRAGE through overall ranking performance, geometric
rectification, component analysis, long-tail recommendation, hyperparameter
sensitivity, and inference efficiency.

\subsection{Experimental Setup}

\subsubsection{Datasets \& Baselines}
To evaluate across diverse topological structures and sparsity levels, we use four standard Amazon review datasets (Beauty, Sports, Toys, and CDs)~\cite{he2016ups}.
\footnote{The datasets are available at \url{https://jmcauley.ucsd.edu/data/amazon/}.}
Following established protocols~\cite{kang2018sasrec,sun2019bert4rec}, we apply
5-core filtering and retain each user's 20 most recent interactions 
(statistics in Table~\ref{tab:dataset_statistics}).

We evaluate MIRAGE against three groups of baselines. 
\textbf{Traditional models} include GRU4Rec~\cite{hidasi2016gru4rec} based on recurrent networks, Caser~\cite{tang2018caser} and NextItNet~\cite{yuan2019nextitnet} using convolutional networks, HGN~\cite{ma2019hgn} with hierarchical gates, graph-based LightGCN~\cite{he2020lightgcn}, and attention-based SASRec~\cite{kang2018sasrec} and BERT4Rec~\cite{sun2019bert4rec}.
\textbf{Diffusion-based models} consist of ADRec \cite{chen2025unlocking} for staged token-level diffusion, DiffuRec \cite{li2024diffurec} for distributional target generation, DimeRec \cite{li2025dimerec} for spherical geodesic diffusion, TA-Rec \cite{mao2025tarec} for one-step denoising, and DreamRec \cite{yang2023dreamrec} for oracle-like target reconstruction.
\textbf{Flow-matching models} include FMRec~\cite{liu2025fmrec}, which constructs straight continuous trajectories and applies deterministic Euler solvers for direct target recovery, and FAVE~\cite{shi2026fave}, which jointly leverages a history-informed semantic prior and regularizes flow consistency to enable accurate one-step generation.

\subsubsection{Metrics \& Implementation Details}
We adopt leave-one-out evaluation with full-catalog ranking and report Hit Rate (H@K) and Normalized Discounted Cumulative Gain (N@K) for
$K\in\{10,20\}$~\cite{he2020lightgcn}.
All experiments are conducted on an NVIDIA RTX 4090 GPU. The flow network $f_\theta$ utilizes a Transformer with 4 blocks, 4 heads, and a hidden dimension $d=128$, while the history decoder $g_\omega$ consists of 2 layers with widths 512 and 2048. We optimize using Adam with a learning rate of $10^{-3}$ and a batch size of 512, applying early stopping governed by validation N@20. 
We fix $\lambda_{\text{hist}}=0.4$~\cite{liu2025fmrec}, $\epsilon=0.001$, tuning other parameters via validation and detailing sensitivity in Appendix~\ref{app:hyperparameter-details}.
Hard alignment proves optimal for Beauty, Sports, and Toys, whereas CDs require the soft variant. Inference strictly defaults to direct one-step generation with $q=1$. 
To ensure robustness, our main performance comparisons report averages across five random seeds (2022–2026). 
All subsequent analyses fix seed 2026 to isolate individual experimental factors.

\subsection{Performance and Geometric Rectification}
\label{subsec:overall_rectification}

\begin{table}[t!]
    \small
    \setlength{\tabcolsep}{1.8pt} 
    \centering
    \caption{Statistics of the datasets.}
    \label{tab:dataset_statistics}
        \begin{tabular}{lrrrrr}
            \toprule
            \textbf{Dataset} & \textbf{\#Users} & \textbf{\#Items} & \textbf{\#Interactions} & \textbf{Avg. Len.} & \textbf{Sparsity} \\
            \midrule
            Beauty & 22,363 & 12,101 & 198,502 & 8.88 & 99.93\% \\
            Sports & 35,598 & 18,357 & 296,337 & 8.32 & 99.95\% \\
            Toys   & 19,412 & 11,924 & 167,597 & 8.63 & 99.93\% \\
            CDs    & 75,258 & 64,443 & 1,097,592 & 14.58 & 99.98\% \\
            \bottomrule
        \end{tabular}
\end{table}

\providecommand{\best}[1]{\textbf{#1}}
\providecommand{\second}[1]{\underline{#1}}
\definecolor{smokypurple}{HTML}{EAE4F2}
\definecolor{stdcolor}{HTML}{666666}

\begin{table*}[t]
\centering
\caption{Main results on four datasets (\%; higher is better). \textbf{Best} and \underline{second} are marked. $\pm$ denotes standard deviation over 5 runs. MIRAGE significantly outperforms the best baselines (paired t-test, $p<0.05$).}
\vspace{-3pt}
\label{tab:main-comparison}
\setlength{\tabcolsep}{1.5pt}
\resizebox{\textwidth}{!}{%
\begin{tabular}{clcccccccccccccccc}
\toprule
& & \multicolumn{4}{c}{Beauty} & \multicolumn{4}{c}{Sports} & \multicolumn{4}{c}{Toys} & \multicolumn{4}{c}{CDs} \\
\cmidrule(lr){3-6} \cmidrule(lr){7-10} \cmidrule(lr){11-14} \cmidrule(lr){15-18}
\multirow{-2}{*}{\textbf{Type}} & \multirow{-2}{*}{\textbf{Model}} & H@10 & H@20 & N@10 & N@20 & H@10 & H@20 & N@10 & N@20 & H@10 & H@20 & N@10 & N@20 & H@10 & H@20 & N@10 & N@20 \\
\midrule
\multirow{7}{*}{\rotatebox[origin=c]{90}{\textit{Trad.}}} & GRU4Rec & 2.53\textcolor{stdcolor}{{\tiny $\pm$0.10}} & 4.26\textcolor{stdcolor}{{\tiny $\pm$0.22}} & 1.21\textcolor{stdcolor}{{\tiny $\pm$0.05}} & 1.64\textcolor{stdcolor}{{\tiny $\pm$0.10}} & 1.92\textcolor{stdcolor}{{\tiny $\pm$0.12}} & 2.96\textcolor{stdcolor}{{\tiny $\pm$0.18}} & 1.01\textcolor{stdcolor}{{\tiny $\pm$0.07}} & 1.27\textcolor{stdcolor}{{\tiny $\pm$0.07}} & 1.79\textcolor{stdcolor}{{\tiny $\pm$0.09}} & 3.19\textcolor{stdcolor}{{\tiny $\pm$0.17}} & 0.86\textcolor{stdcolor}{{\tiny $\pm$0.07}} & 1.21\textcolor{stdcolor}{{\tiny $\pm$0.06}} & 3.77\textcolor{stdcolor}{{\tiny $\pm$0.25}} & 6.29\textcolor{stdcolor}{{\tiny $\pm$0.37}} & 1.86\textcolor{stdcolor}{{\tiny $\pm$0.10}} & 2.50\textcolor{stdcolor}{{\tiny $\pm$0.12}} \\
& Caser & 2.37\textcolor{stdcolor}{{\tiny $\pm$0.11}} & 3.97\textcolor{stdcolor}{{\tiny $\pm$0.25}} & 1.16\textcolor{stdcolor}{{\tiny $\pm$0.07}} & 1.56\textcolor{stdcolor}{{\tiny $\pm$0.10}} & 1.82\textcolor{stdcolor}{{\tiny $\pm$0.10}} & 2.78\textcolor{stdcolor}{{\tiny $\pm$0.15}} & 0.90\textcolor{stdcolor}{{\tiny $\pm$0.05}} & 1.14\textcolor{stdcolor}{{\tiny $\pm$0.08}} & 1.75\textcolor{stdcolor}{{\tiny $\pm$0.11}} & 2.74\textcolor{stdcolor}{{\tiny $\pm$0.12}} & 0.88\textcolor{stdcolor}{{\tiny $\pm$0.07}} & 1.13\textcolor{stdcolor}{{\tiny $\pm$0.06}} & 3.00\textcolor{stdcolor}{{\tiny $\pm$0.12}} & 5.00\textcolor{stdcolor}{{\tiny $\pm$0.18}} & 1.48\textcolor{stdcolor}{{\tiny $\pm$0.10}} & 1.98\textcolor{stdcolor}{{\tiny $\pm$0.12}} \\
& HGN & 1.94\textcolor{stdcolor}{{\tiny $\pm$0.10}} & 3.09\textcolor{stdcolor}{{\tiny $\pm$0.20}} & 0.93\textcolor{stdcolor}{{\tiny $\pm$0.06}} & 1.22\textcolor{stdcolor}{{\tiny $\pm$0.10}} & 1.22\textcolor{stdcolor}{{\tiny $\pm$0.08}} & 2.02\textcolor{stdcolor}{{\tiny $\pm$0.12}} & 0.67\textcolor{stdcolor}{{\tiny $\pm$0.06}} & 0.87\textcolor{stdcolor}{{\tiny $\pm$0.07}} & 1.32\textcolor{stdcolor}{{\tiny $\pm$0.07}} & 2.27\textcolor{stdcolor}{{\tiny $\pm$0.10}} & 0.68\textcolor{stdcolor}{{\tiny $\pm$0.06}} & 0.92\textcolor{stdcolor}{{\tiny $\pm$0.06}} & 0.57\textcolor{stdcolor}{{\tiny $\pm$0.05}} & 0.96\textcolor{stdcolor}{{\tiny $\pm$0.06}} & 0.29\textcolor{stdcolor}{{\tiny $\pm$0.04}} & 0.39\textcolor{stdcolor}{{\tiny $\pm$0.04}} \\
& NextItNet & 4.47\textcolor{stdcolor}{{\tiny $\pm$0.28}} & 7.14\textcolor{stdcolor}{{\tiny $\pm$0.31}} & 2.20\textcolor{stdcolor}{{\tiny $\pm$0.10}} & 2.87\textcolor{stdcolor}{{\tiny $\pm$0.17}} & 2.65\textcolor{stdcolor}{{\tiny $\pm$0.14}} & 4.33\textcolor{stdcolor}{{\tiny $\pm$0.28}} & 1.35\textcolor{stdcolor}{{\tiny $\pm$0.07}} & 1.77\textcolor{stdcolor}{{\tiny $\pm$0.09}} & 3.24\textcolor{stdcolor}{{\tiny $\pm$0.13}} & 5.09\textcolor{stdcolor}{{\tiny $\pm$0.23}} & 1.62\textcolor{stdcolor}{{\tiny $\pm$0.06}} & 2.08\textcolor{stdcolor}{{\tiny $\pm$0.13}} & 5.45\textcolor{stdcolor}{{\tiny $\pm$0.20}} & 8.56\textcolor{stdcolor}{{\tiny $\pm$0.34}} & 2.73\textcolor{stdcolor}{{\tiny $\pm$0.17}} & 3.51\textcolor{stdcolor}{{\tiny $\pm$0.20}} \\
& LightGCN & 4.38\textcolor{stdcolor}{{\tiny $\pm$0.17}} & 6.90\textcolor{stdcolor}{{\tiny $\pm$0.35}} & 2.12\textcolor{stdcolor}{{\tiny $\pm$0.14}} & 2.76\textcolor{stdcolor}{{\tiny $\pm$0.16}} & 2.79\textcolor{stdcolor}{{\tiny $\pm$0.12}} & 4.73\textcolor{stdcolor}{{\tiny $\pm$0.31}} & 1.43\textcolor{stdcolor}{{\tiny $\pm$0.08}} & 1.92\textcolor{stdcolor}{{\tiny $\pm$0.12}} & 4.30\textcolor{stdcolor}{{\tiny $\pm$0.26}} & 6.33\textcolor{stdcolor}{{\tiny $\pm$0.27}} & 2.14\textcolor{stdcolor}{{\tiny $\pm$0.14}} & 2.65\textcolor{stdcolor}{{\tiny $\pm$0.11}} & 5.18\textcolor{stdcolor}{{\tiny $\pm$0.20}} & 7.98\textcolor{stdcolor}{{\tiny $\pm$0.27}} & 2.62\textcolor{stdcolor}{{\tiny $\pm$0.13}} & 3.32\textcolor{stdcolor}{{\tiny $\pm$0.12}} \\
& SASRec & 4.89\textcolor{stdcolor}{{\tiny $\pm$0.21}} & 7.69\textcolor{stdcolor}{{\tiny $\pm$0.32}} & 2.11\textcolor{stdcolor}{{\tiny $\pm$0.14}} & 2.82\textcolor{stdcolor}{{\tiny $\pm$0.17}} & 2.95\textcolor{stdcolor}{{\tiny $\pm$0.15}} & 4.71\textcolor{stdcolor}{{\tiny $\pm$0.21}} & 1.26\textcolor{stdcolor}{{\tiny $\pm$0.07}} & 1.70\textcolor{stdcolor}{{\tiny $\pm$0.09}} & 5.67\textcolor{stdcolor}{{\tiny $\pm$0.34}} & 8.31\textcolor{stdcolor}{{\tiny $\pm$0.41}} & 2.47\textcolor{stdcolor}{{\tiny $\pm$0.13}} & 3.13\textcolor{stdcolor}{{\tiny $\pm$0.14}} & 4.79\textcolor{stdcolor}{{\tiny $\pm$0.28}} & 7.90\textcolor{stdcolor}{{\tiny $\pm$0.39}} & 2.08\textcolor{stdcolor}{{\tiny $\pm$0.12}} & 2.86\textcolor{stdcolor}{{\tiny $\pm$0.13}} \\
& BERT4Rec & 4.13\textcolor{stdcolor}{{\tiny $\pm$0.19}} & 6.27\textcolor{stdcolor}{{\tiny $\pm$0.28}} & 2.20\textcolor{stdcolor}{{\tiny $\pm$0.10}} & 2.74\textcolor{stdcolor}{{\tiny $\pm$0.12}} & 2.03\textcolor{stdcolor}{{\tiny $\pm$0.13}} & 3.48\textcolor{stdcolor}{{\tiny $\pm$0.17}} & 1.01\textcolor{stdcolor}{{\tiny $\pm$0.06}} & 1.37\textcolor{stdcolor}{{\tiny $\pm$0.10}} & 3.54\textcolor{stdcolor}{{\tiny $\pm$0.22}} & 5.18\textcolor{stdcolor}{{\tiny $\pm$0.22}} & 1.86\textcolor{stdcolor}{{\tiny $\pm$0.13}} & 2.27\textcolor{stdcolor}{{\tiny $\pm$0.12}} & 5.66\textcolor{stdcolor}{{\tiny $\pm$0.34}} & 8.70\textcolor{stdcolor}{{\tiny $\pm$0.53}} & 2.85\textcolor{stdcolor}{{\tiny $\pm$0.12}} & 3.61\textcolor{stdcolor}{{\tiny $\pm$0.19}} \\
\midrule
\multirow{5}{*}{\rotatebox[origin=c]{90}{\textit{Diff.}}} & ADRec & 7.62\textcolor{stdcolor}{{\tiny $\pm$0.12}} & 10.52\textcolor{stdcolor}{{\tiny $\pm$0.09}} & 4.49\textcolor{stdcolor}{{\tiny $\pm$0.05}} & 5.22\textcolor{stdcolor}{{\tiny $\pm$0.06}} & 4.03\textcolor{stdcolor}{{\tiny $\pm$0.14}} & 5.96\textcolor{stdcolor}{{\tiny $\pm$0.13}} & 2.21\textcolor{stdcolor}{{\tiny $\pm$0.06}} & 2.69\textcolor{stdcolor}{{\tiny $\pm$0.05}} & 7.75\textcolor{stdcolor}{{\tiny $\pm$0.20}} & 10.37\textcolor{stdcolor}{{\tiny $\pm$0.25}} & 4.84\textcolor{stdcolor}{{\tiny $\pm$0.07}} & 5.50\textcolor{stdcolor}{{\tiny $\pm$0.08}} & 6.86\textcolor{stdcolor}{{\tiny $\pm$0.17}} & 9.91\textcolor{stdcolor}{{\tiny $\pm$0.22}} & 3.69\textcolor{stdcolor}{{\tiny $\pm$0.07}} & 4.46\textcolor{stdcolor}{{\tiny $\pm$0.08}} \\
& DiffuRec & 7.63\textcolor{stdcolor}{{\tiny $\pm$0.08}} & 10.49\textcolor{stdcolor}{{\tiny $\pm$0.15}} & 4.58\textcolor{stdcolor}{{\tiny $\pm$0.09}} & 5.30\textcolor{stdcolor}{{\tiny $\pm$0.11}} & 3.70\textcolor{stdcolor}{{\tiny $\pm$0.23}} & 5.51\textcolor{stdcolor}{{\tiny $\pm$0.20}} & 2.03\textcolor{stdcolor}{{\tiny $\pm$0.19}} & 2.49\textcolor{stdcolor}{{\tiny $\pm$0.18}} & 7.26\textcolor{stdcolor}{{\tiny $\pm$0.14}} & 9.74\textcolor{stdcolor}{{\tiny $\pm$0.18}} & 4.59\textcolor{stdcolor}{{\tiny $\pm$0.05}} & 5.21\textcolor{stdcolor}{{\tiny $\pm$0.06}} & 8.55\textcolor{stdcolor}{{\tiny $\pm$0.50}} & 12.55\textcolor{stdcolor}{{\tiny $\pm$0.64}} & 4.41\textcolor{stdcolor}{{\tiny $\pm$0.25}} & 5.41\textcolor{stdcolor}{{\tiny $\pm$0.28}} \\
& DimeRec & 6.25\textcolor{stdcolor}{{\tiny $\pm$0.10}} & 9.76\textcolor{stdcolor}{{\tiny $\pm$0.18}} & 3.07\textcolor{stdcolor}{{\tiny $\pm$0.10}} & 3.96\textcolor{stdcolor}{{\tiny $\pm$0.11}} & 3.70\textcolor{stdcolor}{{\tiny $\pm$0.11}} & 5.89\textcolor{stdcolor}{{\tiny $\pm$0.13}} & 1.93\textcolor{stdcolor}{{\tiny $\pm$0.08}} & 2.48\textcolor{stdcolor}{{\tiny $\pm$0.08}} & 5.86\textcolor{stdcolor}{{\tiny $\pm$0.27}} & 8.86\textcolor{stdcolor}{{\tiny $\pm$0.31}} & 2.91\textcolor{stdcolor}{{\tiny $\pm$0.12}} & 3.66\textcolor{stdcolor}{{\tiny $\pm$0.12}} & 5.52\textcolor{stdcolor}{{\tiny $\pm$0.16}} & 8.85\textcolor{stdcolor}{{\tiny $\pm$0.26}} & 2.76\textcolor{stdcolor}{{\tiny $\pm$0.09}} & 3.60\textcolor{stdcolor}{{\tiny $\pm$0.11}} \\
& TA-Rec & 0.42\textcolor{stdcolor}{{\tiny $\pm$0.04}} & 0.85\textcolor{stdcolor}{{\tiny $\pm$0.06}} & 0.21\textcolor{stdcolor}{{\tiny $\pm$0.02}} & 0.35\textcolor{stdcolor}{{\tiny $\pm$0.03}} & 0.38\textcolor{stdcolor}{{\tiny $\pm$0.03}} & 0.72\textcolor{stdcolor}{{\tiny $\pm$0.06}} & 0.18\textcolor{stdcolor}{{\tiny $\pm$0.03}} & 0.25\textcolor{stdcolor}{{\tiny $\pm$0.03}} & 0.55\textcolor{stdcolor}{{\tiny $\pm$0.05}} & 0.92\textcolor{stdcolor}{{\tiny $\pm$0.06}} & 0.25\textcolor{stdcolor}{{\tiny $\pm$0.03}} & 0.38\textcolor{stdcolor}{{\tiny $\pm$0.03}} & 0.48\textcolor{stdcolor}{{\tiny $\pm$0.05}} & 0.81\textcolor{stdcolor}{{\tiny $\pm$0.06}} & 0.22\textcolor{stdcolor}{{\tiny $\pm$0.02}} & 0.31\textcolor{stdcolor}{{\tiny $\pm$0.02}} \\
& DreamRec & 0.05\textcolor{stdcolor}{{\tiny $\pm$0.02}} & 0.08\textcolor{stdcolor}{{\tiny $\pm$0.02}} & 0.02\textcolor{stdcolor}{{\tiny $\pm$0.02}} & 0.03\textcolor{stdcolor}{{\tiny $\pm$0.02}} & 0.03\textcolor{stdcolor}{{\tiny $\pm$0.02}} & 0.06\textcolor{stdcolor}{{\tiny $\pm$0.02}} & 0.01\textcolor{stdcolor}{{\tiny $\pm$0.03}} & 0.02\textcolor{stdcolor}{{\tiny $\pm$0.03}} & 0.04\textcolor{stdcolor}{{\tiny $\pm$0.03}} & 0.09\textcolor{stdcolor}{{\tiny $\pm$0.03}} & 0.02\textcolor{stdcolor}{{\tiny $\pm$0.03}} & 0.03\textcolor{stdcolor}{{\tiny $\pm$0.01}} & 0.06\textcolor{stdcolor}{{\tiny $\pm$0.02}} & 0.12\textcolor{stdcolor}{{\tiny $\pm$0.03}} & 0.03\textcolor{stdcolor}{{\tiny $\pm$0.03}} & 0.04\textcolor{stdcolor}{{\tiny $\pm$0.03}} \\
\midrule
& FMRec & \second{8.20}\textcolor{stdcolor}{{\tiny $\pm$0.10}} & \second{11.47}\textcolor{stdcolor}{{\tiny $\pm$0.16}} & \second{4.79}\textcolor{stdcolor}{{\tiny $\pm$0.04}} & \second{5.61}\textcolor{stdcolor}{{\tiny $\pm$0.04}} & \second{4.36}\textcolor{stdcolor}{{\tiny $\pm$0.23}} & \second{6.45}\textcolor{stdcolor}{{\tiny $\pm$0.23}} & \second{2.43}\textcolor{stdcolor}{{\tiny $\pm$0.19}} & \second{2.95}\textcolor{stdcolor}{{\tiny $\pm$0.19}} & \second{8.31}\textcolor{stdcolor}{{\tiny $\pm$0.08}} & \second{11.21}\textcolor{stdcolor}{{\tiny $\pm$0.14}} & \second{5.17}\textcolor{stdcolor}{{\tiny $\pm$0.04}} & \second{5.90}\textcolor{stdcolor}{{\tiny $\pm$0.05}} & \second{8.77}\textcolor{stdcolor}{{\tiny $\pm$0.23}} & \second{12.97}\textcolor{stdcolor}{{\tiny $\pm$0.32}} & \second{4.49}\textcolor{stdcolor}{{\tiny $\pm$0.11}} & \second{5.55}\textcolor{stdcolor}{{\tiny $\pm$0.13}} \\
& FAVE & 7.97\textcolor{stdcolor}{{\tiny $\pm$0.27}} & 11.44\textcolor{stdcolor}{{\tiny $\pm$0.39}} & 4.20\textcolor{stdcolor}{{\tiny $\pm$0.13}} & 5.08\textcolor{stdcolor}{{\tiny $\pm$0.16}} & 2.90\textcolor{stdcolor}{{\tiny $\pm$1.80}} & 4.52\textcolor{stdcolor}{{\tiny $\pm$2.49}} & 1.56\textcolor{stdcolor}{{\tiny $\pm$0.95}} & 1.96\textcolor{stdcolor}{{\tiny $\pm$1.13}} & 5.35\textcolor{stdcolor}{{\tiny $\pm$4.01}} & 7.53\textcolor{stdcolor}{{\tiny $\pm$5.34}} & 2.80\textcolor{stdcolor}{{\tiny $\pm$2.07}} & 3.35\textcolor{stdcolor}{{\tiny $\pm$2.40}} & 8.21\textcolor{stdcolor}{{\tiny $\pm$0.23}} & 12.34\textcolor{stdcolor}{{\tiny $\pm$0.30}} & 4.14\textcolor{stdcolor}{{\tiny $\pm$0.07}} & 5.18\textcolor{stdcolor}{{\tiny $\pm$0.09}} \\
\rowcolor{smokypurple} \cellcolor{white} & MIRAGE & \best{8.57}\textcolor{stdcolor}{{\tiny $\pm$0.12}} & \best{12.32}\textcolor{stdcolor}{{\tiny $\pm$0.09}} & \best{4.95}\textcolor{stdcolor}{{\tiny $\pm$0.08}} & \best{5.89}\textcolor{stdcolor}{{\tiny $\pm$0.11}} & \best{4.76}\textcolor{stdcolor}{{\tiny $\pm$0.11}} & \best{7.13}\textcolor{stdcolor}{{\tiny $\pm$0.10}} & \best{2.52}\textcolor{stdcolor}{{\tiny $\pm$0.09}} & \best{3.12}\textcolor{stdcolor}{{\tiny $\pm$0.07}} & \best{8.93}\textcolor{stdcolor}{{\tiny $\pm$0.29}} & \best{12.48}\textcolor{stdcolor}{{\tiny $\pm$0.21}} & \best{5.33}\textcolor{stdcolor}{{\tiny $\pm$0.24}} & \best{6.22}\textcolor{stdcolor}{{\tiny $\pm$0.19}} & \best{9.10}\textcolor{stdcolor}{{\tiny $\pm$0.17}} & \best{13.52}\textcolor{stdcolor}{{\tiny $\pm$0.25}} & \best{4.64}\textcolor{stdcolor}{{\tiny $\pm$0.13}} & \best{5.75}\textcolor{stdcolor}{{\tiny $\pm$0.16}} \\
\rowcolor{smokypurple} \multirow{-4}{*}{\rotatebox[origin=c]{90}{\textit{Flow}}} \cellcolor{white} & Imp.(\%) & +4.51 & +7.41 & +3.34 & +4.99 & +9.17 & +10.54 & +3.70 & +5.76 & +7.46 & +11.33 & +3.09 & +5.42 & +3.76 & +4.24 & +3.34 & +3.60 \\
\bottomrule
\end{tabular}%
}
\vspace{-3pt}
\end{table*}

\subsubsection{Overall Performance}

Table~\ref{tab:main-comparison} summarizes the recommendation performance, where MIRAGE achieves the best results across all datasets and metrics. Compared to the strongest baseline FMRec, our framework delivers substantial relative improvements, highlighted by a 10.54\% and 11.33\% increase in H@20 on the Sports and Toys datasets. 
Conversely, DreamRec and TA-Rec experience significant performance degradation. Under our rigorous full-catalog protocol, DreamRec's omission of negative sampling \cite{chen2025unlocking, li2025dimerec} and TA-Rec's preference alignment on sparse sequences lead to reduced embedding dispersion, making target items difficult to distinguish.
While continuous generative approaches generally outperform traditional models, MIRAGE maintains a distinct advantage over standard flow-based methods. This advantage validates our core structural hypothesis that optimizing for a straight trajectory leaves the intermediate states of standard methods vulnerable to unsupported continuous regions. By dynamically rectifying the embedding geometry during training to enforce continuous semantic support, MIRAGE prevents generated representations from drifting into the Euclidean void, thereby translating straight transport routes into accurate next-item predictions.

\subsubsection{Geometric Rectification}

\begin{figure}[htbp]
    \centering
    \includegraphics[width=\columnwidth]{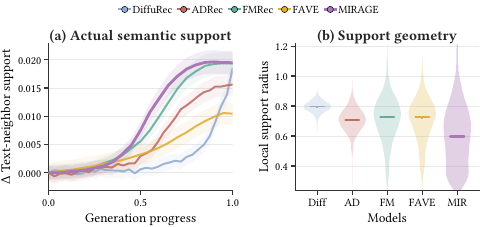}
    \caption{Geometric rectification analysis. (a) Semantic support along the actual generation trajectories, evaluated using an independent text embedding space. (b) Distributions of the normalized local support radius across the item catalog.}
    \Description{A 1x2 figure. Panel (a) shows the relative gain in text-neighbor support over generation progress; MIRAGE rises earlier and higher than baselines. Panel (b) shows violin plots where MIRAGE achieves the lowest median normalized local support radius.}
    \label{fig:geometric_rectification}
\end{figure}

To evaluate whether models escape the Euclidean void during inference, we analyze the intermediate states along the actual generated trajectories. Specifically, we retrieve the nearest neighbors of these states within the learned embedding space and measure their similarity to the true target using a frozen, independent text embedding space.\footnote{Semantic similarity is computed using cosine similarity on frozen item text embeddings extracted via Sentence-T5.} Figure~\ref{fig:geometric_rectification}(a) illustrates the relative gain in this text-neighbor support. While baselines traverse regions with sparse target semantics, MIRAGE exhibits an earlier and more pronounced rise in semantic alignment.
This confirms our training proxy anchors actual paths, a trend that directly correlates with the overall accuracy improvements in Table~\ref{tab:main-comparison}.

Beyond securing individual trajectories, we examine the global catalog geometry in Figure~\ref{fig:geometric_rectification}(b) using the local support radius, defined as the normalized distance from each item to its neighbors. The violin plots show that MIRAGE shifts this distribution downward, yielding a lower median radius than the baselines. A smaller normalized radius reflects denser semantic clustering across the catalog. This indicates that our manifold-informed regularization reorganizes the underlying Euclidean space, establishing a more compact and cohesive environment for continuous generation.

\subsection{Ablation Study}
\label{subsec:ablation}

\providecommand{\best}[1]{\textbf{#1}}
\definecolor{smokypurple}{HTML}{EAE4F2} 

\begin{table}[htbp]
\centering
\small
\setlength{\tabcolsep}{1.8pt} 
\caption{Ablation study of MIRAGE.}
\label{tab:mirage-ablation}
{
\begin{tabular}{lcccccccc}
\toprule
\multicolumn{1}{l}{\multirow{2}{*}{Variant}} & \multicolumn{2}{c}{Beauty} & \multicolumn{2}{c}{Sports} & \multicolumn{2}{c}{Toys} & \multicolumn{2}{c}{CDs} \\
\cmidrule(lr){2-3} \cmidrule(lr){4-5} \cmidrule(lr){6-7} \cmidrule(lr){8-9}
 & H@20 & N@20 & H@20 & N@20 & H@20 & N@20 & H@20 & N@20 \\
\midrule
\rowcolor{smokypurple} MIRAGE & \best{12.27} & \best{5.71} & \best{7.22} & \best{3.12} & \best{12.78} & \best{6.03} & \best{14.17} & \best{6.02} \\
{\scriptsize w/o Topology} & 11.67 & 5.64 & 6.57 & 3.04 & 10.99 & 5.82 & 13.54 & 5.85 \\
{\scriptsize w/o Time Mod.} & 12.25 & 5.63 & 6.55 & 2.82 & 12.74 & 5.90 & 13.63 & 5.78 \\
{\scriptsize w/ Alt. Alignment}  & 11.98 & 5.45 & 5.67 & 2.44 & 12.66 & 5.98 & 13.75 & 5.77 \\
{\scriptsize w/ Endpoint Reg.}  & 10.16 & 4.47 & 5.25 & 2.18 & 10.12 & 4.59 & 13.69 & 5.74 \\
\bottomrule
\end{tabular}
}
\end{table}


To disentangle the contributions of MIRAGE, Table~\ref{tab:mirage-ablation} compares the full MIRAGE framework against four specific variants. 
The ablated variants are defined as follows: 
1) \textbf{w/o Topology} disables the geometric regularizer entirely by setting the weight $\lambda_{\text{topo}}=0$ in Eq.~\eqref{eq:method_total_loss}; 
2) \textbf{w/o Time Mod.} removes the parabolic schedule $w(t)=4t(1-t)$ defined in Eq.~\eqref{eq:method_time_weight}, uniformly applying a constant topology penalty across the entire interpolation timeframe; 
3) \textbf{w/ Alt. Alignment} inverts the optimal dataset-specific configuration by switching between the hard discrete anchor $\boldsymbol{e}_{j_t^*}$ and a softmax-weighted distance;
and 4) \textbf{w/ Endpoint Reg.} acts as static endpoint regularization by minimizing the L2 distance between target and neighbor embeddings, isolating our path-interior gains.

Removing the topology term yields substantial performance degradation, highlighted by a drop in H@20 from $12.78$ to $10.99$ on the Toys dataset. This confirms that the generative flow backbone alone, despite its sequence modeling capabilities, struggles with the Euclidean void and relies on explicit geometric regularization to maintain valid semantic support. Furthermore, replacing the time modulation schedule with a constant penalty undermines recommendation quality, reducing H@20 from $7.22$ to $6.55$ on Sports. This outcome validates that applying the topological constraint indiscriminately near the boundary endpoints ($t \to 0$ or $1$) corrupts the unstructured source noise and the clean target representations.

Switching alignment configurations degrades performance across all domains. Dense datasets rely on discrete hard anchors to keep intermediate states out of unsupported regions, while sparse datasets like CDs require continuous soft aggregation to stabilize local gradients. 
This demonstrates that alignment techniques must naturally adapt to the underlying structural sparsity. 
Furthermore, static endpoint regularization at the target endpoint performs worse than omitting the graph entirely. Forcing target items too close to their neighbors collapses their distinct representations and degrades precise ranking. This confirms that semantic support must be applied dynamically to the intermediate states along the trajectory.

\subsection{Long-Tail Performance}
\label{subsec:long_tail}

To evaluate performance across item popularities, we divide the catalog based on training-set frequency. 
Items in the bottom 80\% of the catalog ranked by their training frequency form the long tail, while the remaining 20\% constitute the head.

\begin{figure}[t!]
    \centering
    \includegraphics[width=\columnwidth]{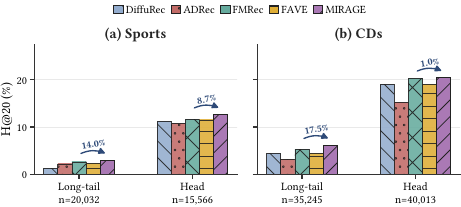}
    \caption{H@20 for Long-Tail and Head Items on Sports/CDs ($n$ denotes the number of test targets in each group).}
    \Description{Bar charts compare long-tail and head H@20; MIRAGE has its
    largest relative gains on long-tail targets.}
    \label{fig:long-tail}
\end{figure}

As shown in Figure~\ref{fig:long-tail}, MIRAGE achieves its most substantial relative gains on long-tail targets. On Sports, it improves H@20 over the strongest baseline by 14.0\% for the tail compared to 8.7\% for the head. On CDs, these respective gains reach 17.5\% and 1.0\%, with consistent trends across the remaining datasets detailed in Appendix~\ref{app:additional-long-tail}. This pronounced improvement on sparse items reflects the utility of our target-local support. Infrequent items receive fewer gradient updates during training, leaving their surrounding embedding geometry poorly defined by standard target supervision. 
By leveraging shared co-occurrence anchors, MIRAGE allows these sparse targets to draw geometric support from high-frequency related items, accumulating local evidence throughout the shared embedding space. 
Without introducing popularity-specific parameters or sacrificing head accuracy, this mechanism effectively concentrates geometric guidance exactly where the continuous generative trajectory is most vulnerable to the Euclidean void.
Crucially, confining this support to the path interior prevents sparse items from collapsing into popular neighbors, avoiding popularity bias while preserving their distinct rankability.

\subsection{Hyperparameter Analysis}
\label{subsec:hyperparameter}

\begin{figure}[t!]
    \centering
    \includegraphics[width=0.90\columnwidth]{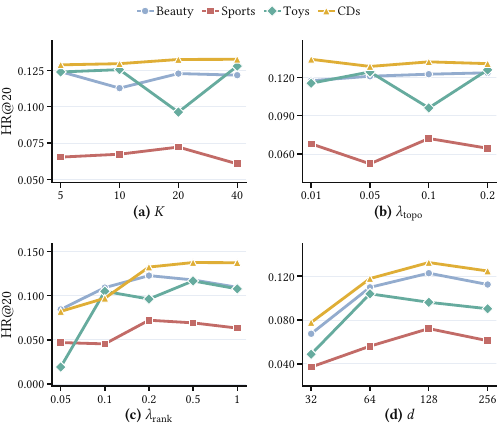}
    \caption{H@20 under one-at-a-time hyperparameter sweeps.}
    \Description{Four line charts vary neighborhood size, topology weight,
    ranking weight, and hidden dimension on four datasets.}
    \label{fig:hyperparameter}
\end{figure}

To systematically assess model sensitivity, Figure~\ref{fig:hyperparameter} shows that optimal neighborhood sizes and loss weights vary across datasets and thus justify our selection through validation. While MIRAGE maintains robust performance over a broad middle range of values, the irregular response on the Toys dataset to both $K$ and $\lambda_{\text{topo}}$ reveals a genuine sensitivity to local graph structure. The neighborhood size governs the balance between support and specificity; a small $K$ provides precise but potentially unstable anchors, whereas a large $K$ increases geometric coverage at the risk of pulling the generative trajectory toward weakly related items. 

Similarly, the interior optima for the topology weight confirm that geometric rectification complements item discrimination. Sweeps over the ranking weight and embedding dimension (peaking at $d=128$) verify that adequate spatial separation and controlled model capacity remain necessary alongside geometric reorganization. Appendix~\ref{app:hyperparameter-details} details the exact search grids and further analyzes the co-occurrence window size $c$, showing that dense catalogs benefit from broader semantic contexts ($c=5$) whereas sparse datasets require strict adjacency ($c=1$) to prevent noisy guidance.

\subsection{Efficiency Analysis}
\label{subsec:efficiency}

\begin{figure}[h]
    \centering
    \includegraphics[width=\columnwidth]{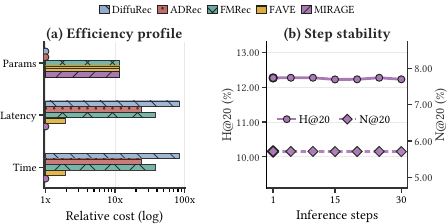}
    \caption{Efficiency profile and step stability. MIRAGE minimizes relative inference latency while maintaining robust accuracy under direct one-step generation.}
    \Description{Left panel shows MIRAGE achieves the lowest relative latency and time among baselines despite matching the parameter count of FMRec. Right panel shows ranking metrics remain stable across 1 to 30 inference steps.}
    \label{fig:inference-steps}
\end{figure}

\begin{figure*}[t!]
    \centering
    \includegraphics[width=0.95\textwidth]{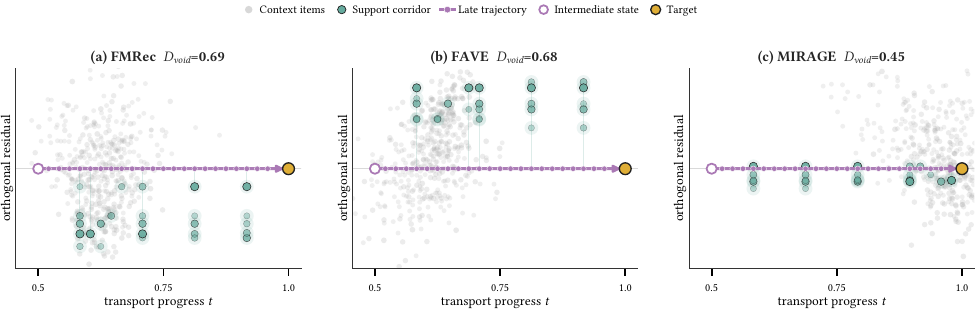}
    \caption{Target-local transport plane visualization. MIRAGE rectifies the embedding space to form a tight support corridor along the late generative trajectory, whereas baselines traverse the unsupported Euclidean void.}
    \Description{A three-panel figure comparing the transport planes of FMRec, FAVE, and MIRAGE. The horizontal axis tracks transport progress while the vertical axis shows the orthogonal residual. For FMRec and FAVE, the nearest support items appear far from the generative trajectory. For MIRAGE, the support items closely hug the trajectory line to form a tight and continuous corridor.}
    \label{fig:trajectory-plane}
\end{figure*}

The efficiency profile in Figure~\ref{fig:inference-steps}(a) compares the relative inference costs of MIRAGE against recent generative recommenders. 
MIRAGE establishes the lowest latency and total inference time, operating orders of magnitude faster than iterative models like DiffuRec while further outperforming the one-step FAVE by eliminating its prior-construction overhead.
Since MIRAGE shares the exact same model capacity as FMRec and FAVE, its superior inference speed stems from its one-step generation strategy rather than a reduced model size.
The topology regularizer and neighborhood retrieval operations are discarded after optimization to achieve zero graph overhead during deployment, while establishing an optimal training time trade-off detailed in Appendix~\ref{app:training-efficiency}.

The step stability plot in Figure~\ref{fig:inference-steps}(b) further justifies this direct generation strategy by tracking recommendation accuracy across varying inference steps. Both metrics remain stable as the integration steps increase from 1 to 30. This stability aligns with our target recovery parameterization and straight probability path. Additional sampling steps only refine a target estimate that already receives direct supervision during training. 
This verifies that our training-time manifold rectification shapes a well-behaved continuous vector field, enabling safe and accurate one-step traversal.
These extra steps therefore multiply the computational cost without supplying new semantic evidence or measurable accuracy gains and confirm one-step inference as the optimal practical setting.

\subsection{Qualitative Trajectory Support Analysis}
\label{app:trajectory-visualization}

To illustrate how MIRAGE reshapes the latent geometry, Figure~\ref{fig:trajectory-plane} visualizes the target-local transport plane during the late generation phase. The horizontal axis represents the interpolant's progress toward the target, while the vertical axis measures the orthogonal residual deviation from the straight probability path. For sampled time bins along the trajectory, faint lines connect the continuous path to its nearest valid graph anchors in the embedding space.

Generative baselines like FMRec and FAVE project straight paths that traverse sparse continuous regions. Their nearest valid item representations exhibit large orthogonal residuals and uneven distributions, yielding long projection lines that indicate a lack of local semantic grounding. By contrast, MIRAGE reorganizes the underlying item geometry to envelop these straight generative trajectories. The nearest graph anchors tightly align with the trajectory and distribute along the generation progress. These aligned anchors form a continuous support corridor that provides dense structural support for the intermediate states toward the intended target semantics.

\section{Conclusion and Future Work}
\label{sec:conclusion}

In this paper, we identify the Euclidean void as a critical failure mode in continuous generative recommendation. 
We demonstrate that straight interpolation paths often traverse unsupported regions and cause intermediate states to lose semantic guidance. 
To address this limitation, we propose MIRAGE, which dynamically rectifies the underlying item geometry to ensure continuous semantic support throughout the generation process. 
This reorganization prevents representation drift while preserving an efficient inference mechanism. 
Extensive evaluations on four datasets confirm that this geometric grounding delivers robust performance improvements and yields particular benefits for sparsely observed items.

While MIRAGE uses Euclidean space, future work will explore hyperbolic embeddings for hierarchical item relationships. These manifolds capture complex category structures and asymmetric user interactions. Incorporating such geometric priors into the transport process could provide even tighter bounds on intermediate representations, further improving generative reliability.

\begin{acks}
This work was partially supported by the
\grantsponsor{nkrd}{National Key Research and Development Program of China}{}
(No.~\grantnum{nkrd}{2023YFF0905400}),
the \grantsponsor{nsfc}{National Natural Science Foundation of China}{https://doi.org/10.13039/501100001809}
(Nos.~\grantnum{nsfc}{U2341229},
\grantnum{nsfc}{62406125}, and
\grantnum{nsfc}{62502404}), the \grantsponsor{moe}{Fundamental and Interdisciplinary Disciplines Breakthrough Plan of the
Ministry of Education of China}{}
(No.~\grantnum{moe}{JYB2025XDXM903}),
the \grantsponsor{rgc}{Research Grants Council of Hong Kong}{}
(Research Impact Fund No.~\grantnum{rgc}{R1015-23};
Collaborative Research Fund No.~\grantnum{rgc}{C1043-24GF};
RGC Research Fellow Scheme No.~\grantnum{rgc}{RFS2627-1S03};
General Research Fund Nos.~\grantnum{rgc}{11218325},
\grantnum{rgc}{11212926}),
the \grantsponsor{jilin}{Reform Commission Foundation of Jilin Province}{}
(No.~\grantnum{jilin}{2024C003}),
and the \grantsponsor{idm}{Institute of Digital Medicine of City University of Hong Kong}{}
(No.~\grantnum{idm}{9229503}).
We also acknowledge support from the
\grantsponsor{ncsf}{New Cornerstone Science Foundation}{}
through the XPLORER PRIZE, as well as industry funding from
\grantsponsor{huawei}{Huawei}{}
(Huawei Innovation Research Program),
\grantsponsor{bytedance}{ByteDance}{},
\grantsponsor{tencent}{Tencent}{}
(CCF-Tencent Open Fund,
Tencent Rhino-Bird Focused Research Program,
Tencent University Cooperation Project),
\grantsponsor{kuaishou}{Kuaishou}{}
(CCF-Kuaishou Large Model Explorer Fund
No.~\grantnum{kuaishou}{2025008},
Kuaishou University Cooperation Project),
\grantsponsor{alibaba}{Alibaba}{}
(CCF-Taobao \& Tmall Group TECH Kangaroo Fund
No.~\grantnum{alibaba}{2026001},
Alibaba Innovative Research Program),
and \grantsponsor{didi}{Didi}{}
(CCF-DiDi GAIA Collaborative Research Funds).
\end{acks}

\clearpage
\bibliographystyle{ACM-Reference-Format}
\bibliography{main}

\clearpage
\appendix
\section{Detailed Methodology}
\label{app:detailed-methodology}

\subsection{Alignment Derivations}
\label{app:alignment-derivation}

\subsubsection{Hard Alignment}

For a target item $i^+$, we define the squared Euclidean distance between the interpolated state $\boldsymbol{x}_t$ and a graph neighbor as follows
\begin{equation}
    \phi_j(\boldsymbol{x}_t)
    =\left\|\boldsymbol{x}_t-\boldsymbol{e}_j\right\|_2^2,
    \quad j\in\mathcal{N}_K(i^+).
\end{equation}
When the closest neighbor $j_t^*$ is unique, it remains the nearest anchor within a sufficiently small neighborhood of $\boldsymbol{x}_t$. The distance function and its gradient are therefore given locally by
\begin{equation}
\begin{aligned}
d_{\mathcal{G}}(\boldsymbol{x}_t,i^+)
&=\phi_{j_t^*}(\boldsymbol{x}_t),\\
\nabla_{\boldsymbol{x}_t}d_{\mathcal{G}}(\boldsymbol{x}_t,i^+)
&=2\left(\boldsymbol{x}_t-\boldsymbol{e}_{j_t^*}\right).
\end{aligned}
\label{eq:app-hard-gradient}
\end{equation}
Thus, hard alignment directly guides the interpolated state toward its nearest valid item embedding.

\subsubsection{Soft Alignment}

Soft alignment assigns temperature-controlled weights to
neighboring item distances, with $\tau>0$:
\begin{equation}
\begin{aligned}
d_{\mathcal{G},\tau}^{\text{soft}}(\boldsymbol{x}_t,i^+)
&=\sum_{j\in\mathcal{N}_K(i^+)}
\alpha_{t,j}\phi_j(\boldsymbol{x}_t),\\
\alpha_{t,j}
&=\frac{\exp(-\phi_j(\boldsymbol{x}_t)/\tau)}
{\sum_{a\in\mathcal{N}_K(i^+)}
\exp(-\phi_a(\boldsymbol{x}_t)/\tau)}.
\end{aligned}
\label{eq:app-soft-alignment}
\end{equation}

Differentiating the weighted distance also accounts for
the dependence of $\alpha_{t,j}$ on $\boldsymbol{x}_t$:
\begin{equation}
\begin{aligned}
&\nabla_{\boldsymbol{x}_t}
d_{\mathcal{G},\tau}^{\text{soft}}\\
&= \sum_{j\in\mathcal{N}_K(i^+)}
\alpha_{t,j}
\left(
1-\frac{\phi_j-d_{\mathcal{G},\tau}^{\text{soft}}}{\tau}
\right)
\nabla_{\boldsymbol{x}_t}\phi_j\\
&=2\sum_{j\in\mathcal{N}_K(i^+)}
\alpha_{t,j}
\left(
1-\frac{\phi_j-d_{\mathcal{G},\tau}^{\text{soft}}}{\tau}
\right)
(\boldsymbol{x}_t-\boldsymbol{e}_j),
\end{aligned}
\label{eq:app-soft-gradient}
\end{equation}
where $\phi_j=\phi_j(\boldsymbol{x}_t)$ for brevity.
The gradient adapts each neighbor's contribution according
to its distance and alignment weight.

As $\tau\rightarrow 0$, the weight of a closest neighbor $j_t^*$ approaches 1. With the distance gap
$\Delta_j=\phi_j(\boldsymbol{x}_t)-\phi_{j_t^*}(\boldsymbol{x}_t)>0$
for $j\neq j_t^*$, we obtain
\begin{equation}
\alpha_{t,j_t^*}=
\left(
1+\sum_{j\neq j_t^*}
\exp\left(-\frac{\Delta_j}{\tau}\right)
\right)^{-1}
\longrightarrow 1
\quad\text{as}\quad \tau\rightarrow 0.
\label{eq:app-soft-limit}
\end{equation}
Meanwhile, the weights of all other neighbors approach zero. Therefore, soft alignment and its gradient converge to their hard-alignment counterparts wherever the closest neighbor is unique.

\subsection{Gradient Dynamics}
\label{app:gradient-dynamics}

To understand how the topological loss reshapes the embedding space, we analyze the gradient dynamics during a local optimization step. Assuming a training state where the active anchor $j_t^*$ remains locally constant and is distinct from the target item, we define the residual vector as $\boldsymbol{r}=\boldsymbol{x}_t-\boldsymbol{e}_{j_t^*}$. Applying a gradient step of size $\eta$ updates the target and anchor embeddings according to:
\begin{equation}
\begin{aligned}
\boldsymbol{e}_{i^+}^{\prime}
&=\boldsymbol{e}_{i^+}-2\eta w(t)t\boldsymbol{r},\\
\boldsymbol{e}_{j_t^*}^{\prime}
&=\boldsymbol{e}_{j_t^*}+2\eta w(t)\boldsymbol{r}.
\end{aligned}
\label{eq:app-embedding-update}
\end{equation}
Since the interpolation is defined as $\boldsymbol{x}_t=(1-t)\boldsymbol{x}_0+t\boldsymbol{e}_{i^+}$, this update to the target embedding simultaneously shifts the intermediate state to $\boldsymbol{x}_t^{\prime}=\boldsymbol{x}_t-2\eta w(t)t^2\boldsymbol{r}$. Consequently, the new residual becomes:
\begin{equation}
\boldsymbol{r}^{\prime}
=\boldsymbol{x}_t^{\prime}-\boldsymbol{e}_{j_t^*}^{\prime}
=\left[1-2\eta w(t)(1+t^2)\right]\boldsymbol{r}.
\label{eq:app-residual-contraction}
\end{equation}
A learning rate constrained within $0<\eta w(t)(1+t^2)<1$ ensures this local step decreases the residual norm. Even with additional gradients from the recovery, ranking, and history components during full training, the topological term enforces this local contraction toward valid semantic anchors. This dynamic accounts for the implicit geometric rectification.

\subsection{Inference Consistency}
\label{app:inference-consistency}

To demonstrate the efficiency of single-step generation where $q=1$, we evaluate the Euler integration over the full interval length $1-\epsilon$. Initializing the integration with $\tilde{\boldsymbol{x}}_0=\boldsymbol{x}_0$ at $t_0=\epsilon$ and substituting the parameterized velocity yields:
\begin{equation}
\begin{aligned}
\tilde{\boldsymbol{x}}_1
&=\boldsymbol{x}_0+(1-\epsilon)\left[f_\theta(\boldsymbol{x}_0,\epsilon,S_u)-\boldsymbol{x}_0\right]\\
&=(1-\epsilon)f_\theta(\boldsymbol{x}_0,\epsilon,S_u)+\epsilon\boldsymbol{x}_0.
\end{aligned}
\label{eq:app-onestep-reduction}
\end{equation}
Since the boundary margin $\epsilon$ is chosen to be negligibly small, this expression naturally reduces to the direct target prediction formulation $\hat{\boldsymbol{x}}_1$. Furthermore, because the co-occurrence graph is utilized exclusively within the topological loss during training, it remains decoupled from the learned vector field and inference equations, eliminating the need for graph lookups during deployment.

\section{Additional Experimental Details}

\subsection{Baseline Implementation Details}
\label{app:baseline-implementation}

All evaluated baselines were strictly reproduced using their official open-source implementations and recommended protocols. Recent related models, such as FlowRec and DAE4Rec, are omitted because their official source codes are not publicly available.

\subsection{Hyperparameter Sensitivity}
\label{app:hyperparameter-details}

\begin{figure}[htbp]
    \centering
    \includegraphics[width=\columnwidth]{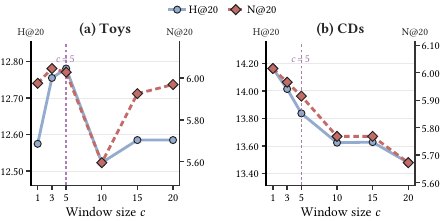}
    \caption{Effect of the co-occurrence window size $c$. }
    \Description{A 1x2 figure displaying line plots of H at 20 and N at 20 across window sizes 1, 3, 5, 10, 15, and 20. The left panel for Toys shows both metrics peaking at c=5 to form an inverted U-shape. The right panel for CDs shows both metrics starting at their highest point at c=1 and steadily decreasing as the window size increases.}
    \label{fig:window_size}
\end{figure}

We evaluate hyperparameter sensitivity by isolating individual variables. The optimal neighborhood size $K \in \{5, 10, 20, 40\}$ depends on data sparsity. Sparser domains like CDs and Toys benefit from broader structural support with $K=40$, whereas the denser Beauty dataset peaks at $K=5$. Furthermore, performance exhibits pronounced dataset-specific sensitivity to both the topology weight $\lambda_{\text{topo}}$ and the ranking weight $\lambda_{\text{rank}}$. Toys, in particular, experience degradation if these weights deviate from their respective optima of $0.2$ and $0.5$, underscoring the necessity of balancing geometric rectification with the contrastive ranking objective.

As illustrated in Figure~\ref{fig:window_size}, evaluating the co-occurrence window size $c \in \{1, 3, 5, 10, 15, 20\}$ reveals distinct geometric requirements across different data distributions. Dense catalogs like Toys achieve optimal performance at an intermediate window of $c=5$, forming a pronounced inverted U-shaped trend. This shows that moderately larger windows capture broader semantic contexts to support hard alignment before eventually introducing overly smoothed correlations. Conversely, the sparse and sequential CDs dataset favors adjacent co-occurrences at $c=1$. Extending the window for this sparse catalog rapidly introduces distant noisy edges, which dilute the weighted centroid of the soft alignment variant and consequently degrade the geometric guidance.

\subsection{Graph Construction and Complexity}

We construct the topology graph once using the training split exclusively. 
By concatenating the history and target for each training instance, we count ordered item pairs co-occurring within a dataset-specific context window ($c=1$ for CDs and $c=5$ otherwise).
The stored neighborhood for every item comprises those with the highest positive co-occurrence frequencies, resolving ties deterministically to ensure reproducibility. Because this graph remains fixed during optimization, dynamic embedding updates never alter neighborhood memberships or leak future information.

Processing a mini-batch of size $B$ with hard alignment requires retrieving up to $K$ neighbor embeddings per target to compute their squared distances to the interpolated state. This introduces a distance computation cost of $\mathcal{O}(BKd)$ and caches $\mathcal{O}(|\mathcal{I}|K)$ indices. The soft alignment applies a normalized weighting over these distances, resulting in identical asymptotic complexity. Both strategies reuse the existing item table without introducing auxiliary encoders. Furthermore, we discard the cached graph and neighbor retrieval mechanisms after training, which preserves the baseline parameter count and ensures zero graph overhead during inference.

\subsection{Training Efficiency}
\label{app:training-efficiency}

\begin{figure}[htbp]
    \centering
    \includegraphics[width=\columnwidth]{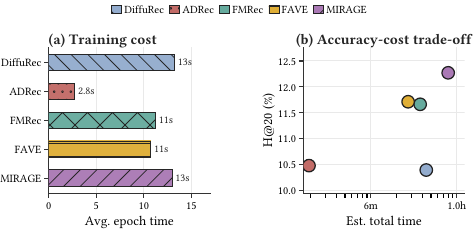}
    \caption{Training efficiency and performance trade-off on Beauty. MIRAGE requires comparable per-epoch training time while establishing the optimal Pareto frontier in overall accuracy and total training cost.}
    \Description{A 1x2 figure. The left panel is a bar chart comparing average epoch time, where MIRAGE and DiffuRec take 13 seconds, FMRec and FAVE take 11 seconds, and ADRec takes 2.8 seconds. The right panel is a scatter plot of Hit Rate at 20 versus estimated total time, showing MIRAGE at the highest accuracy with a total time comparable to DiffuRec and FMRec.}
    \label{fig:training-efficiency}
\end{figure}

To address potential concerns regarding the computational overhead of topological regularization, Figure~\ref{fig:training-efficiency} compares the average epoch time and the overall accuracy-efficiency trade-off. MIRAGE requires $13$ seconds per epoch, remaining comparable to standard generative baselines like FMRec and FAVE which require 11 seconds. The scatter plot confirms that MIRAGE achieves a favorable Pareto trade-off by yielding the highest performance for a similar overall training budget. This demonstrates that the geometric rectification process introduces negligible computational burden while delivering recommendation improvements.

\subsection{Additional Long-Tail Results}
\label{app:additional-long-tail}

\begin{figure}[htbp]
    \centering
    \includegraphics[width=\columnwidth]{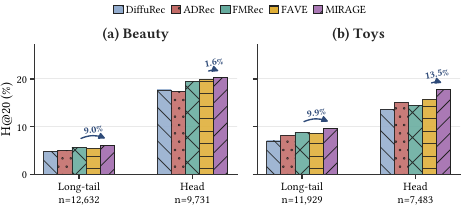}
    \caption{Additional long-tail H@20 results on Beauty and Toys. Buckets use training-set item frequency ($n$ denotes the number of test targets in each group).}
    \Description{Side-by-side bar charts of H at 20 for long-tail and head targets on Beauty and Toys. MIRAGE is compared with DiffuRec, ADRec, FMRec, and FAVE.}
    \label{fig:long-tail-appendix}
\end{figure}
To evaluate long-tail performance, we partition the bottom $80\%$ of the catalog based on training-set frequency as the long tail and the remaining $20\%$ as the head. Test instances are assigned to these fixed subsets based on ground-truth targets to ensure fair, model-independent evaluations.
As shown in Figure~\ref{fig:long-tail-appendix}, MIRAGE improves H@20 over the strongest baseline by $9.0\%$ for long-tail targets and $1.6\%$ for head targets on Beauty. Toys achieves respective gains of $9.9\%$ on the tail and $13.5\%$ on the head. Alongside the main text results for Sports and CDs, these findings show that MIRAGE consistently enhances generative recommendations for sparsely observed items without sacrificing accuracy on popular targets.

Unlike other domains, Toys exhibits a gain of $13.5\%$ on head targets alongside its $9.9\%$ tail improvement. This concurrent improvement stems from the unique category characteristics of toys, where popular items share tightly coupled interaction sequences. In such densely correlated item spaces, our topological regularization effectively captures high-frequency structural paths without suffering from performance saturation, simultaneously boosting popular and sparsely observed recommendations.

\end{document}